**Title:** Strain-invariant, highly water stable all-organic soft conductors based on ultralight multi-layered foam-like framework structures

**Authors:**

I. Barg,[1] N. Kohlmann,[2] F. Rasch,[3] T. Strunskus,[1,4] R. Adelung,[3,4] L. Kienle,[2,4] F. Faupel,[1,4] S. Schröder,[1,4]*† F. Schütt[3,4]*†

**Affiliations:**
[1]Chair for Multicomponent Materials, Institute for Materials Science, Kiel University; 24143 Kiel, Germany
[2]Chair for Synthesis and Real Structure, Institute for Materials Science; 24143 Kiel University, Kiel, Germany
[3]Chair for Functional Nanomaterials, Institute for Materials Science, Kiel University; 24143 Kiel, Germany
[4]Kiel Nano, Surface and Interface Science KiNSIS, Kiel University; 24118 Kiel, Germany

*Corresponding authors. Email:
Dr.-Ing. Fabian Schütt (fas@tf.uni-kiel.de)
Dr.-Ing. Stefan Schröder (ssch@tf.uni-kiel.de)

†Shared senior authorship

**Abstract:** Soft and flexible conductors are essential in the development of soft robots, wearable electronics, as well as electronic tissue and implants. However, conventional soft conductors are inherently characterized by a large change in conductance upon mechanical deformation or under alternating environmental conditions, e.g., humidity, drastically limiting their application potential and performance. Here, we demonstrate a novel concept for the development of strain-invariant, fatigue resistant and highly water stable soft conductor. By combining different thin film technologies in a three-dimensional fashion, we develop nano- and micro-engineered, multi-layered (< 50 nm), ultra-lightweight (< 15 mg/cm³) foam-like composite framework structures based on PEDOT:PSS and PTFE. The all-organic composite framework structures are characterized by conductivities of up to 184 S/m, remaining strain-invariant between 80 % compressive and 25 % tensile strain. We further show, that the multi-layered composites are characterized by properties that surpass that of framework structures based on the individual materials. Both, the initial electrical and mechanical properties of the composite framework structures are retained during long-term cycling, even after 2000 cycles at 50 % compression. Furthermore, the PTFE functionalization renders the framework structure highly hydrophobic, resulting in stable electrical properties, even when immersed in water for up to 30 days. The here presented concept overcomes the previous limitations of strain-invariant soft conductors and demonstrates for the first time a versatile approach for the development of innovative multi-scaled and multi-layered functional materials, for applications in soft electronics, energy storage and conversion, sensing, catalysis, water and air purification, as well as biomedicine.

**One-Sentence Summary:** Piezo-resistive framework structures are converted into strain-invariant, highly water stable soft conductors by a nanoscale coating.

**Main Text:**

## INTRODUCTION

Today's electronics, such as smart phones or laptops, are typically based on semiconducting and metallic conductors. However, due to their rigid nature, the application of these conventional electronic components in fields of wearable electronics, soft robotics, as well as electronic implants is limited. In order to develop innovative functional electronic devices, that can naturally interact with soft living matter or improve human-machine interaction, new soft (bio-)electronic materials and components such as actuators, sensors, as well as soft and flexible conductors have to be designed (*1–4*). Conventional approaches for the fabrication of soft and flexible conductors include geometric engineering of non-stretchable conductors into stretchable patterns (*5, 6*), incorporation of conductive fillers, such as carbon nanomaterials, into an elastomer matrix (*7*) or the plastization of conductive polymer (CP) thin films through additives (*8*). Advanced fabrication methods include the use of aerogels and multi-scaled framework structures based on conductive nanomaterials, e.g. Graphene or Carbon Nanotubes (*9–11*), as well as aerogels and foams based on conductive polymers (CPs) (*12–14*). Especially, the latter provides (a) the intrinsic capacity for both compressive and tensile strains without failure, (b) high conductivities at ultra-light weight, and (c) adjustable mechanical compliance. Further benefits are breathability, high specific surface area, improved mass and charge transport and higher availability of active sites (*12*). While in combination these properties make aerogels, foams and framework structures promising candidates for the development of high-performance soft conductors, their electrical properties are extremely susceptible to mechanical deformation or changes in environmental conditions, e.g. humidity. For example, CP foams, framework-structures and aerogels are typically characterized by a piezo-resistive effect, i.e. a change in resistance upon mechanical deformation. This effect is often utilized to create strain and force sensors based on CPs (*15–18*). Furthermore, 3D porous framework structures are prone to fatigue upon cyclic load, which for example commonly manifests as cyclic softening, i.e. the decrease in stress at a given strain upon repeated deformation (*19–21*). And finally, the conductivity of CP foam-like structures and their composites are prone to water absorption, which can be utilized for sensing applications (*22–24*), but makes its electrical properties dependent on ambient humidity.

However, for applications in the fields of soft robotics, wearable electronics and electronic tissue technologies constant electrical and mechanical characteristics of conductors, irrespective of deformation state or environmental conditions, e.g. humidity, are of uttermost importance. For example, a reliable electrical readout, communication, and control of electronic components, such as sensors and actuators, connected or integrated into a single soft matrix, can only be achieved by a strain-invariant response of the conductor, i.e. a constant resistance upon deformation. Furthermore, if the soft conductor is a load-bearing component or is used to drive an actuator in soft robotics, a constant mechanical response without fatigue is required.

In this work we present for the first time a novel approach, in which we combine wet-chemical and dry-chemical thin film technologies in a three-dimensional fashion to develop innovative multi-scaled and dual-layered foam-like composite framework structures. We show, that the as prepared composites are characterized by a set of properties that surpass that of the individual materials, enabling completely new functionalities and thus application scenarios. As a first demonstration of this concept, in this work we focus on the development of multi-scaled framework structures based on Poly(3,4-ethylenedioxythiophene):poly(styrenesulfonate) (PEDOT:PSS) and Polytetrafluoroethylene (PTFE). Utilizing a ceramic template, highly porous (porosity up to 99.6 %) and highly conductive (~184 S/m) PEDOT:PSS framework structures are fabricated by a wet chemical approach, consisting of networked thin films (<50 nm) in the which overcomes the aforementioned limitations of aerogels and foam-like framework structures for soft electronics. The approach is based on the synthesis of multi-scaled framework structures of PEDOT:PSS, which forms a network of interconnected hollow microtubes. This three-dimensional macroscopic thin film structure is further functionalized with an insulating polytetrafluoroethylene PTFE thin film coating of on average ~ 23 ± 10 nm via initiated chemical vapor deposition (iCVD), a vapor based free-radical polymerization technique, that has emerged as a highly efficient method for conformal polymer coatings down to nanoscale thicknesses even in complex 3D geometries (*25–27*). Similar to conventional wire insulation, the applied PTFE thin film coating acts as an insulting layer between the PEDOT:PSS microtubes. Thereby, the previously piezo-resistive PEDOT:PSS framework structure is converted into a strain-invariant soft conductor with constant conductivity between 80 % compressive and 25 % tensile strain, without changing the overall morphology of the framework structure. Both, the initial electrical and mechanical properties are retained during long-term cycling at 50 % compression. Furthermore, the PTFE thin film coating results in an enhancement of the mechanical flexibility and stability of the PEDOT:PSS framework structure. Lastly, the PTFE functionalization greatly reduces water absorption, which enables constant electrical properties even after 30 days immersed in water. Whereas other state-of-the-art strain-invariant conductors may feature one of the previously mentioned characteristics, the here presented approach excels in all of them (see Supplementary Tables S1 to S3), thereby overcoming the aforementioned limitations of aerogels and foam-like framework structures for soft electronics.

## RESULTS

### Synthesis

The schematic synthesis route for the PEDOT:PSS framework structure (Aero-PEDOT:PSS) and its subsequent functionalization via iCVD is shown in Fig. 1A. The synthesis uses highly porous (~ 94 %) sacrificial templates of interconnected ZnO microrods, which have proven to be a suitable pathway for the dry-chemical and wet-chemical assembly of nanomaterials into macroscopic and lightweight 3D networks (*9, 10, 28*). The ZnO templates are prepared by a molding process (Fig. 1A, Step 1), in which the geometry as well as density of the final template can be controlled. Infiltration of an aqueous PEDOT:PSS solution by dripping leads to a homogenous distribution throughout the porous network (Fig. 1B and Supplementary Video S1, further details see Methods). After evaporation of the solvent a thin film of PEDOT:PSS with an average thickness of ~ 16 ± 3 nm remains on the ZnO microrods (Fig. 1A, Step 2, and Supplementary Fig. S1). Freestanding Aero-PEDOT:PSS frameworks are then obtained by wet-chemical etch-removal of the sacrificial ZnO and critical point drying (Fig. 1A, Step 3). With densities between 3.8 and 15.0 mg/cm³ the obtained framework structures are highly porous (>99.8 %) and ultra-lightweight, i.e. readily attach to nearby objects such as tissue paper (Fig. 1C). Furthermore, the template-based approach can be utilized to fabricate framework structures that can be specifically controlled in their macroscopic size and geometry (Fig. 1D).

By controlling the amount of infiltrated PEDOT:PSS during synthesis the density (3.8-15.0 mg/cm³) and the corresponding conductivity (22.9-174.8 S/m) of Aero-PEDOT:PSS can be adjusted (Fig. 1E). As highlighted by the conductivity of PEDOT:PSS-ZnO prior to etching (Fig. 1E9, the etching is bifunctional, since it not only removes the sacrificial ZnO template, but also increases the conductivity by 59 % on average. X-ray photoelectron spectroscopy (XPS) analysis of the ratios between the Sulfur 2p peaks of PEDOT and PSS confirms, that the mechanism is the removal of non-conductive PSS through etching, which reduces the ratio between PSS and PEDOT by ~ 56 % (Inset Fig. 1E, detailed deconvolution see Supplementary Fig. S2) (*29–31*). Finally, the as-prepared Aero-PEDOT:PSS framework structures are functionalized with a thin layer of PTFE via iCVD (Fig. 1A, Step 4, and Fig. 1E). Here, the monomer gas hexafluoropropylene oxide (HPFO) (orange) is thermally decomposed at a hot filament into a trifluoroacetyl fluoride byproduct (white) and difluorocarbene (red). The PTFE thin film is then formed by adsorption and chain-growth of the difluorocarbene on the previously synthesized Aero-PEDOT:PSS (blue). The film growth is furthermore accelerated through addition of a thermally activated initiator perfluorobutanesulfonylfluoride (PFBSF) (not shown). A detailed experimental description of the iCVD process is given in the Methods and elsewhere (*25, 32*).

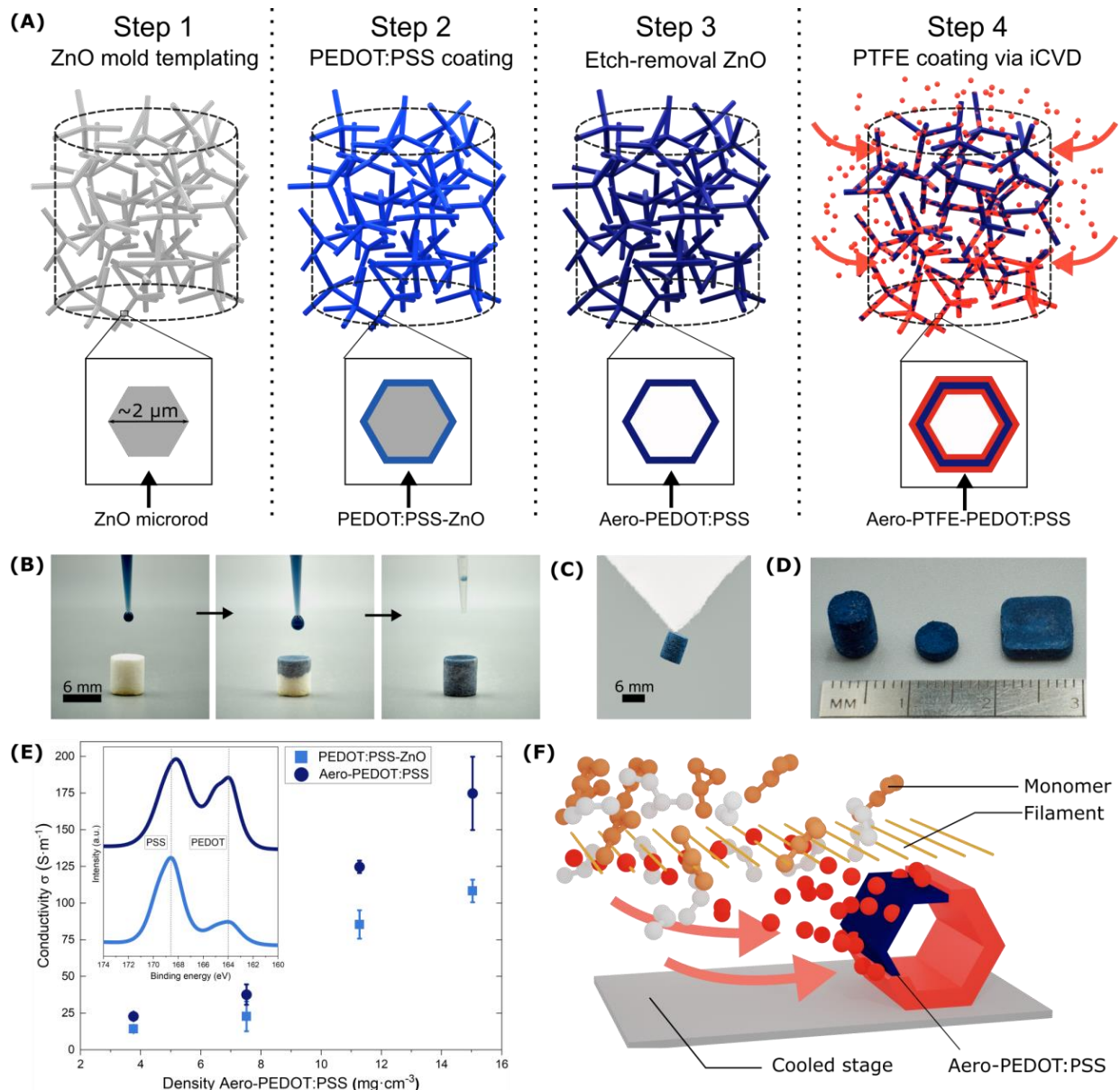

**Fig. 1. Schematic synthesis and conductivity measurement.** (**A**) Schematic depiction of the synthesis process in four steps. (**B**) Photographs showing the infiltration of a t-ZnO template with an aqueous PEDOT:PSS solution. (**C**) Photograph of Aero-PEDOT:PSS hanging from a paper tissue, highlighting the ultralow density of the synthesized material. (**D**) Photograph of resulting Aero-PEDOT:PSS samples with different geometries after etch-removal of ZnO. (**E**) Specific conductivity versus density of Aero-PEDOT:PSS. The corresponding conductivity of samples prior to etching is indicated as PEDOT:PSS-ZnO. Inset: Normalized Sulfur 2p spectra of PSS and PEDOT components for Aero-PEDOT:PSS and PEDOT:PSS-ZnO measured by XPS. The increase in relative height of the PEDOT peak at 164 eV indicates removal of PSS through etching. (**F**) Schematic illustration of the iCVD coating process of an Aero-PEDOT:PSS arm (not to scale).

**Morphology and chemical characterization**

Scanning electron microscopy (SEM) was used to investigate the resulting microstructure of Aero-PEDOT:PSS and Aero-PTFE-PEDOT:PSS. The non-functionalized Aero-PEDOT:PSS (Fig. 2A) consists of a network of interconnected hollow microtubes resembling the tetrapodal ZnO structure of the template with partial sail-like formation in between the arms (detailed image overview see Supplementary Fig. S3). The iCVD PTFE deposition does not change the overall morphology of the network. The comparison of individual Aero-PEDOT:PSS arms before and after PTFE coating show no distinct differences in SEM and indicate a thin and conformal coating (Supplementary Fig. S4).

A more detailed investigation of the nanostructure of the PEDOT:PSS films with and without iCVD deposition of PTFE was carried out using transmission electron microscopy (TEM). Due to the highly flexible nature of Aero-PEDOT:PSS and Aero-PTFE-PEDOT:PSS extraction of single microrods suitable for TEM analysis is not feasible. Instead, structures with intact ZnO cores yet identical coatings of PEDOT:PSS and subsequent iCVD PTFE deposition are investigated. The amorphous polymer shell can readily be distinguished from the crystalline ZnO core by mass-thickness as well as diffraction contrast (Fig. 2C,D) in bright field (BF) TEM. Thus, the thickness of the PEDOT:PSS and PTFE-PEDOT:PSS thin films can be determined from the change in gray value within the TEM micrograph. Investigation of numerous microrods yields ~ $16 \pm 3$ nm thickness for the PEDOT:PSS layer. After iCVD coating a clear increase in average polymer shell thickness of ~ $23 \pm 10$ nm is observed. As stated previously, a small amount of PSS is assumed to be removed in the process of ZnO etching. Thus, the thickness in Aero-PEDOT:PSS can be assumed to be slightly less than the value determined here.

Both polymers possess a unique element, i.e., S for PEDOT:PSS and F for PTFE. Therefore, the presence of the polymers can be inferred from the presence of S and F in the shell respectively. Spatially resolved energy dispersive X-ray spectroscopy (EDX) via scanning TEM (STEM) EDX mapping of a microrod coated with both PEDOT:PSS and PTFE (Inset Fig. 2D) clearly shows a shell containing both S and F around the ZnO core with an outer layer containing no S. Accordingly, the successful iCVD deposition of PTFE on PEDOT:PSS is strongly supported. However, the difference in average shell thickness before and after iCVD coating with PTFE is much greater than this apparent PTFE layer thickness of ~ 3 nm. It can thus be concluded that at least partial intermixing of the polymers takes place. In order to unambiguously verify PTFE deposition additional investigation via a complementary method is performed.

To investigate the chemical structure of PEDOT:PSS before and after PTFE functionalization Fourier transformation infrared spectroscopy (FTIR) was used. Figure 2E shows the FTIR spectra of PEDOT:PSS (blue) and PTFE-PEDOT:PSS (red) between 2000 and 500 cm$^{-1}$ (wide spectra see Supplementary Fig. S5). In both cases characteristic absorption peaks of PEDOT:PSS are present. The peaks at 1583 cm$^{-1}$ and 1537 cm$^{-1}$ are related to the C=C stretch in PSS and PEDOT respectively. At 1307 cm$^{-1}$ the C-C stretch and at 1137 cm$^{-1}$ the C-O-C stretch of PEDOT can be seen (all peaks given in Supplementary Table S4) (*33*, *34*). The highlighted area shows the differences introduced by the PTFE deposition. While the peak around 1220 cm$^{-1}$ corresponds to asymmetrical C-F2 stretching, its overlap with the C-O-C absorption band of PEDOT at 1207 cm$^{-1}$ allows no clear identification without deconvolution. However, the unique peak at 1163 cm$^{-1}$ is attributed to symmetrical CF2 stretching of PTFE, thus indicating a successful deposition of PTFE (*35–37*). This is additionally confirmed by XPS, since for Aero-PTFE-PEDOT:PSS typical Fluorine peaks with F KLL at 833.5 and 859.5 eV (Supplementary Fig. S6) and F 1s at 688.9 eV (Supplementary Fig. S7) appear. Furthermore, the C 1s spectrum shows the presence of a C-F2 peak at 292.0 eV attributed to PTFE (Fig. 2F), which is not present in unfunctionalized Aero-PEDOT:PSS (Supplementary Fig. S8) (25, 35). Noticeably, for a PTFE film thickness exceeding the signal depth in XPS (<10 nm) the C 1s spectrum is expected to be dominated by the C-F2 contribution, which is not the case here. A possible explanation is that the porosity created by the etch-removal of PSS allows for local intermixing of PTFE and PEDOT:PSS, thus reducing the overall film thickness in Aero-PTFE-PEDOT:PSS. While the exact microstructure in Aero-PTFE-PEDOT:PSS is yet to be determined, the functionalization nevertheless leads to drastic changes in its physical properties compared to its unfunctionalized counterpart, as will be demonstrated in the following paragraphs.

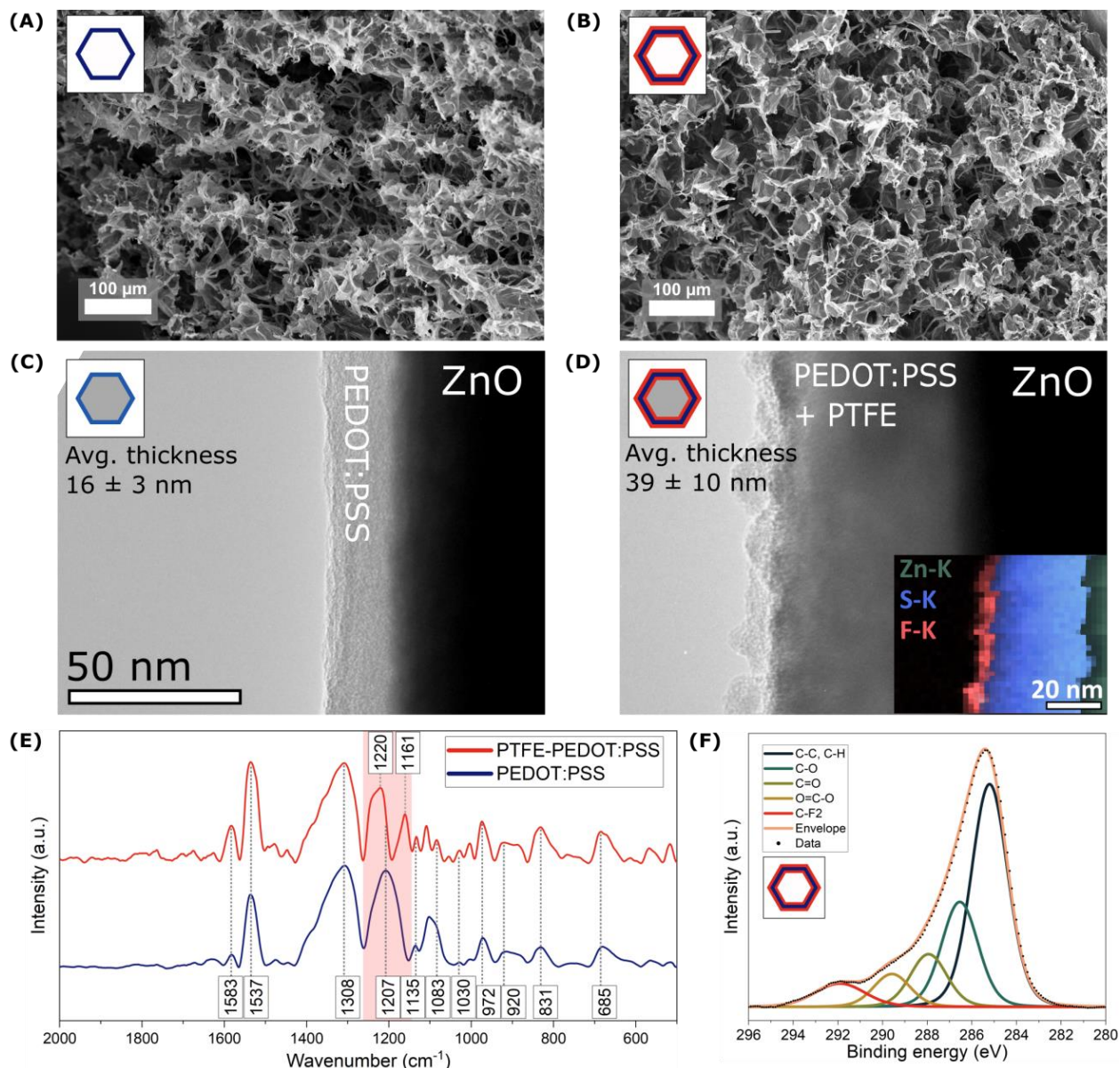

**Fig. 2. Morphology and chemical characterization.** (**A**) SEM micrograph showing a cross-sectional view of Aero-PEDOT:PSS after etch-removal of ZnO. (**B**) SEM micrograph showing a cross-section of Aero-PTFE-PEDOT:PSS. (**C**) BF TEM micrograph of a microrod with intact ZnO core coated by PEDOT:PSS. (**D**) BF TEM micrograph of a similar microrod to (C) after an additional PTFE iCVD coating step. The inset presents a STEM EDX map showing a S and F containing shell around the ZnO core. (**E**) FTIR spectra of PEDOT:PSS and PTFE-PEDOT:PSS. The area highlighted in red denotes changes after PTFE deposition. (**F**) XPS spectrum of Carbon

1s peak for Aero-PTFE-PEDOT:PSS. The additional peak at around 292 eV is attributed to C-F2 binding, confirming a successful deposition of PTFE.

**Strain-invariant conductors**

The influence of the PTFE layer on the mechanical and electrical properties was tested under compressive load using a custom made setup (Supplementary Fig. S9). The stress-strain curves for different compression cycles (40-80 %) for Aero-PEDOT:PSS (Fig. 3A) and Aero-PTFE-PEDOT:PSS (Fig. 3B) both show the typical response of porous 3D networks (*13*, *38*). A minor linear regime (I) is followed by a plateau region (II) and finally a strong non-linear behavior in the densification region (III). Compared to Aero-PEDOT:PSS the maximum stress at 80% compression increases for Aero-PTFE-PEDOT:PSS from 22.7 kPa to 24.1 kPA, while the Young's modulus increases from 0.92 kPA to 7.92 kPa. Because of the high porosity these values are considerably lower than for bulk PEDOT:PSS (9.3 MPa) (*39*) or PTFE deposited by iCVD (34 MPa), making them suitable for tissue compliant electronics. Overall, the PTFE coating gives rise to reversible compression up to 80% with a low residual strain of 2 % compared to 60 % for Aero-PEDOT:PSS (Fig. 3C and Supplementary Video S2). The explanation for this is based on the densification and formation of new interconnections between structural elements of the network upon compression (Fig. 3D). In the absence of a barrier layer, as is the case for Aero-PEDOT:PSS, adhesive forces between the new connections reduce the elastic restoring force and lead to plastic deformation and/or rupture upon unloading. The PTFE encapsulation of the PEDOT:PSS network negates this effect due to the extremely low surface energy of PTFE (*40*). Additionally, PTFE acts as insulating layer due its low dielectric constant (*25*, *40*), which can be seen in the distinct differences in the normalized electrical resistance $R/R_0$ for consecutive compression cycles (40-80%) for Aero-PEDOT:PSS (Fig. 3E) and Aero-PTFE-PEDOT:PSS (Fig. 3F). The non-linear piezoresistive response of Aero-PEDOT:PSS with $R/R_0 = 0.30$ at 80 % compression is caused by the formation of new conductive pathways during compression. A net increase in resistance upon uploading is associated with the reduction of conductive pathways due to residual strain or material failure. Aero-PTFE-PEDOT:PSS on the other hand is highly strain-invariant over a wide range with $R/R_0 = 0.95$ at 80% compression. The stark contrast in behavior is indirect proof for the completeness of PTFE coverage throughout the PEDOT:PSS network. Since a soft conductor component should be able to withstand both compressive and tensile load, maximum tensile strain tests were performed (Fig. 3G). For Aero-PEDOT:PSS an ultimate tensile strength of 0.62 kPa and strain at fracture of 16.1 % can be seen. The stretchability for Aero-PTFE-PEDOT:PSS is improved to a tensile strength of 2.16 kPa and fracture at 47.9 % strain. Aero-PEDOT:PSS is cleary piezoresistive at any given tensile strain, while Aero-PTFE-PEDOT:PSS remains strain-invariant up until 25 % strain with $R/R_0 = 1.04$ (Fig. 3H).

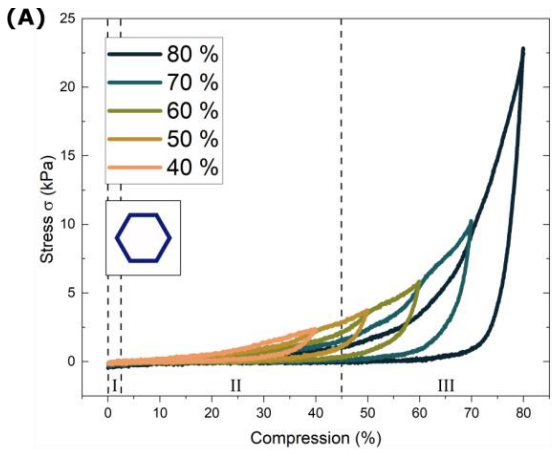
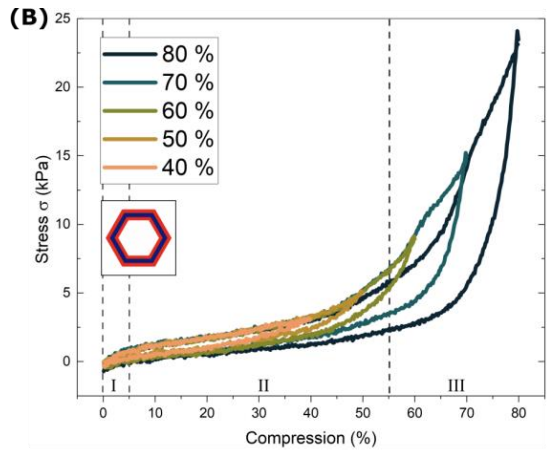
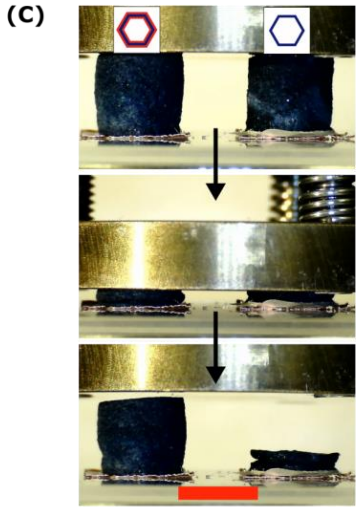
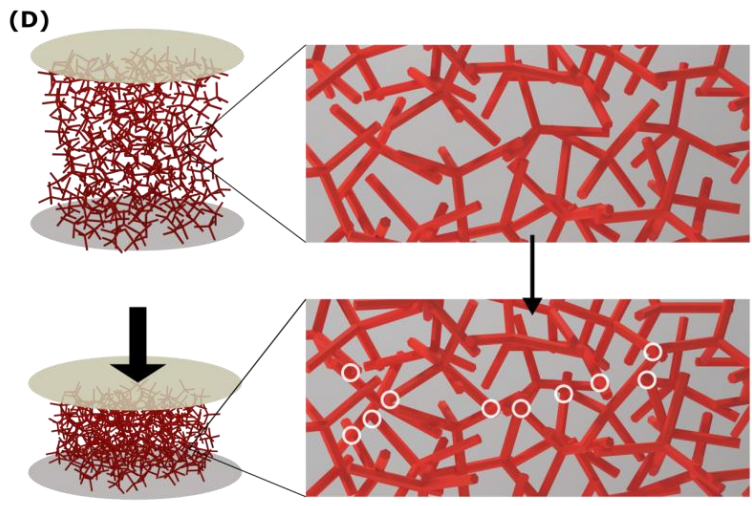
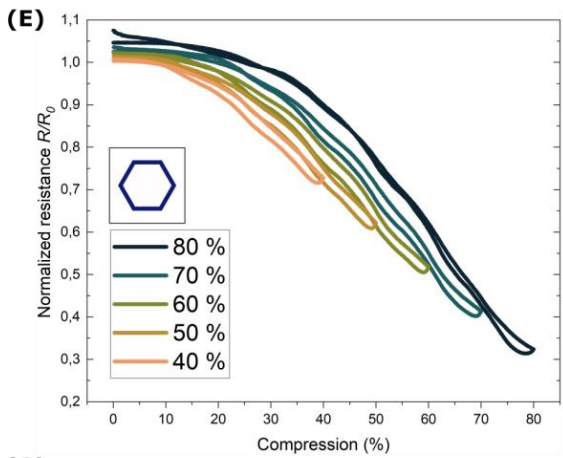
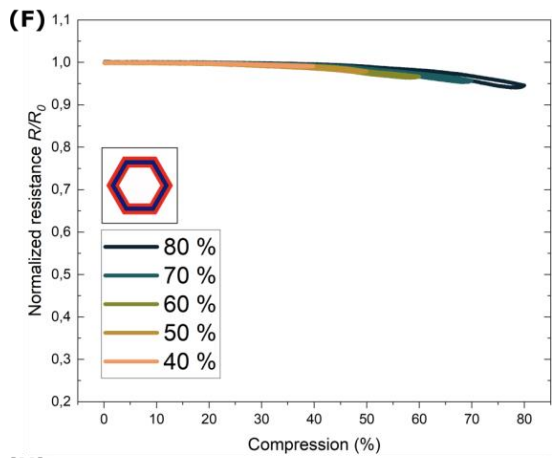
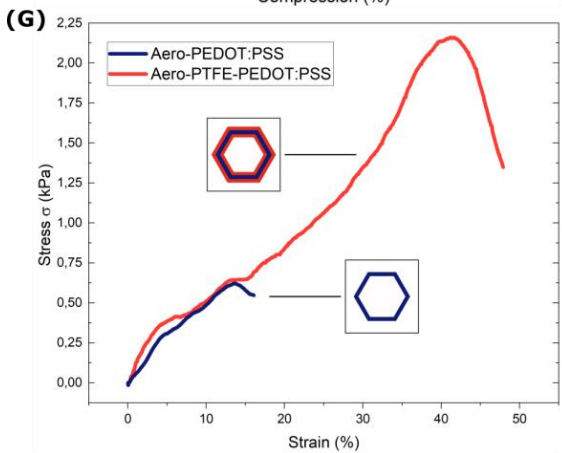
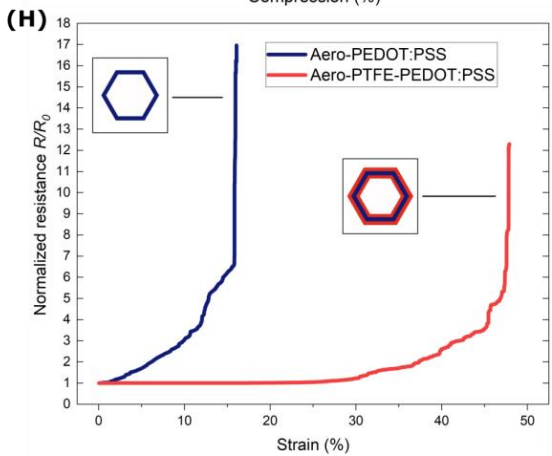

**Fig. 3. Mechanical and electrical characterization.** (**A**) Compressive stress-strain curve between 40 % and 80 % compression for Aero-PEDOT:PSS. (**B**) Compressive stress-strain curve between 40 % and 80 % compression for Aero-PTFE-PEDOT:PSS. (**C**) Photographic series highlighting the response to 80 % compression for Aero-PTFE-PEDOT:PSS (left) and Aero-PEDOT:PSS (right). Scale bar (red) is 6 mm. (**D**) Schematic depiction of compression process and the formation of new interconnections (white circles) upon densification of the network elements. (**E**) Normalized resistance $R/R_0$ versus subsequent compression cycles (40-80%) for Aero-PEDOT:PSS. (**F**) Normalized resistance $R/R_0$ versus subsequent compression cycles (40-80%) for Aero-PTFE-PEDOT:PSS. (**G**) Tensile stress-strain curve until material failure for Aero-PEDOT:PSS and Aero-PTFE-PEDOT:PSS. (**H**) Normalized resistance $R/R_0$ versus tensile strain for Aero-PEDOT:PSS and Aero-PTFE-PEDOT:PSS.

The long-term stability of the materials was tested for 2000 cycles at 50 % compression. The stress-strain curves for the first and last cycle show less residual strain for Aero-PTFE-PEDOT:PSS (4.5 %) compared to Aero-PEDOT:PSS (35.1 %) (Fig. 4A). Aero-PEDOT:PSS is prone to fatigue (Fig. 4B), indicated by a decrease in stress at maximum compression ($\sigma = 3.2$ kPA, $\sigma/\sigma_0 = 0.84$) and a decrease in the energy loss coefficient $Q$, i.e. the area ratio between loading and unloading curve with respect to the loading curve, from 0.63 to 0.30 (calculation see Supplementary Section 9). Aero-PTFE-PEDOT:PSS shows a constant maximum stress ($\sigma_0 = 5.06$ kPa, $\sigma_{end}/\sigma_0 = 0.98$) and a constant loss coefficient ($Q_0 = 0.28$, $Q_{end}/Q_0 = 0.96$). Compared to the literature, Aero-PTFE-PEDOT:PSS shows remarkably low fatigue without any cyclic softening, outperforming any other porous or aeromaterial so far reported (see Supplementary Table S1). The normalized resistance $R/R_0$ versus compression for the 1st and 2000th cycle for both materials is given in Fig. 4C. The piezoresistive Aero-PEDOT:PSS suffers from an continuous increase in resistance at all compression ranges due to ongoing plastic deformation. On the other hand, Aero-PTFE-PEDOT:PSS is highly stable with only minor deviations at maximum compression. While $R/R_0$ at zero compression shows no change, the resistance at 50 % compression changes from 0.98 on the first to 0.95 on the last cycle (Fig. 4D). Compared to the literature, the cyclic stability in its electrical response of Aero-PTFE-PEDOT:PSS matches that of the most stable strain-invariant conductors reported so far (Supplementary Table S2). While Chen *et al.* reported a strain-invariant PEDOT:PSS-BIBSAT foam with similar performance and structure, this approach however suffers from a 4-fold increase in resistance even after 20 minutes in water (*13*).

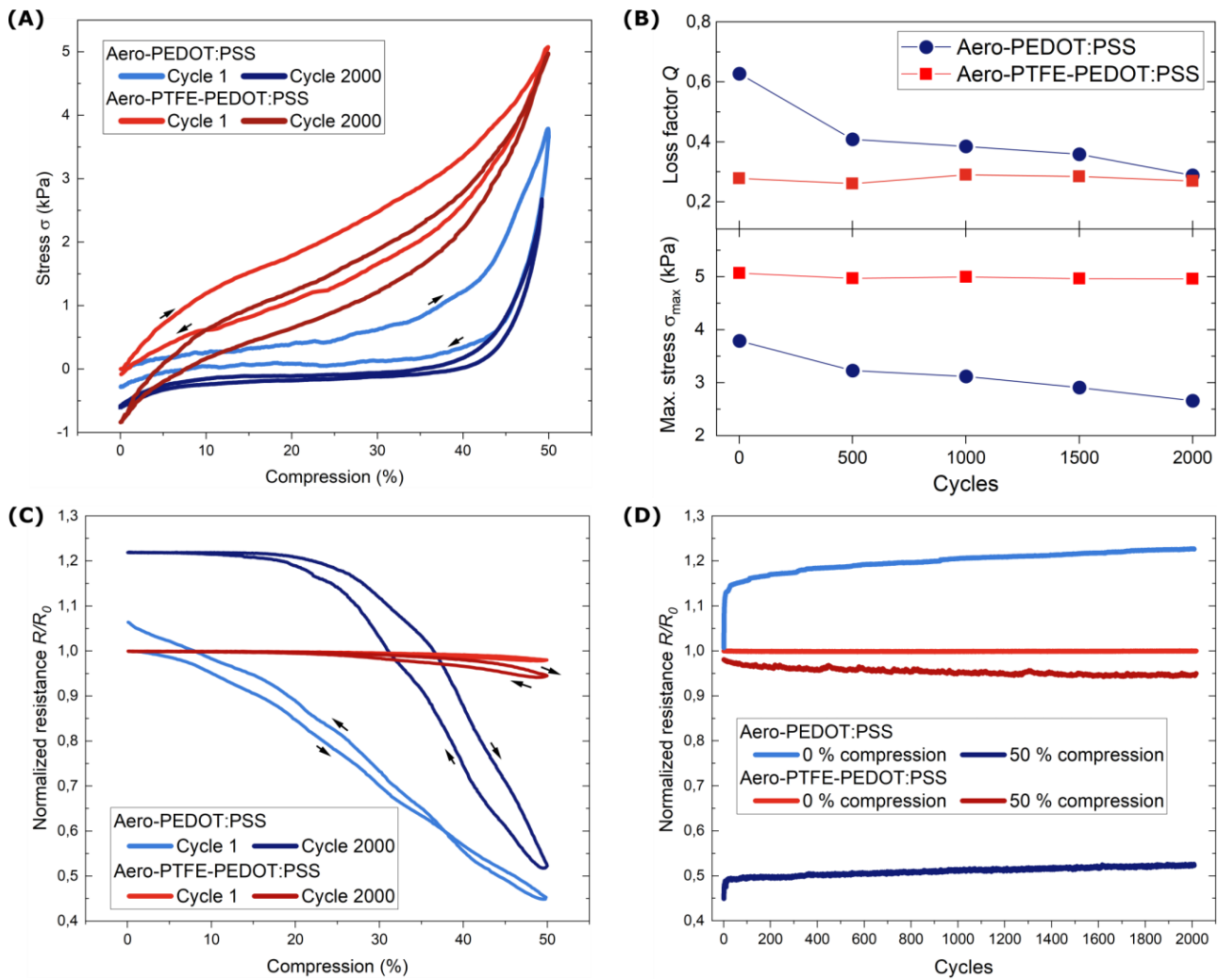

**Fig. 4. Long-term cycling.** Results after 2000 cycles at 50 % compression for Aero-PEDOT:PSS (blue) and Aero-PTFE-PEDOT:PSS (red). (**A**) Compressive stress-strain curve for the 1st and 2000th cycle highlighting the improved elasticity of Aero-PTFE-PEDOT:PSS. (**B**) The energy loss factor and stress at maximum compression (bottom) versus cycle number show a greatly improved stability with little fatigue for Aero-PTFE-PEDOT:PSS. (**C**) Normalized resistance $R/R_0$ versus compression on 1st and 2000th cycle highlighting little to no change in the electrical properties after functionalization. (**D**) The normalized resistance $R/R_0$ at 0 % and 50 % compression versus cycle number shows the highly stable strain-invariant behavior of Aero-PTFE-PEDOT:PSS.

**Water stability**

Contact angle measurements were performed to determine the wetting characteristics before and after PTFE functionalization. In case of Aero-PEDOT:PSS no contact angle can be measured, since the water droplet is immediately absorbed upon contact (Fig. 5A and Supplementary Video S3). On the other hand, Aero-PTFE-PEDOT:PSS is strongly hydrophobic with an average contact angle of about 139 ± 4°, which remains stable over a duration of 10 minutes and longer. To test the effect of the functionalization on the swelling behavior, the swelling ratio $m/m_0$ was measured for different samples for up to 30 days in water (Fig. 5B). As Aero-PTFE-PEDOT:PSS is too hydrophobic to be submerged in water without force (Supplementary Video S4 and S5), the samples were held under water using a sieve. Aero-PEDOT:PSS with a final swelling ratio of 12.94 is prone to quick swelling, which leads to visible bulging of the samples (Fig. 5B, left photograph). On the other hand the strong hydrophobicity of Aero-PTFE-PEDOT:PSS is highlighted by the formation of a distinct air layer underwater and (Fig. 5B, right photograph), and a drastic reduction in water uptake with $m/m_0 = 1.21$ after 30 days. Finally, the electrical characteristics for Aero-PEDOT:PSS and Aero-PTFE-PEDOT:PSS were measured for samples immersed in water for 30 days using a custom made setup (Fig. 5C). After 30 days, the final ratio $R/R_0$ for Aero-PEDOT:PSS is at 2.30 due to water absorption, whereas it is only 1.24 for Aero-PTFE-PEDOT:PSS, highlighting the good protective properties against swelling. Our approach matches the water stability of the best performing other approaches (see Supplementary Table S3), which however either sacrifice the benefits of a 3D porous network by using an elastomeric filler (*41*) or maintain the porosity, but are piezoresistive (*42*).

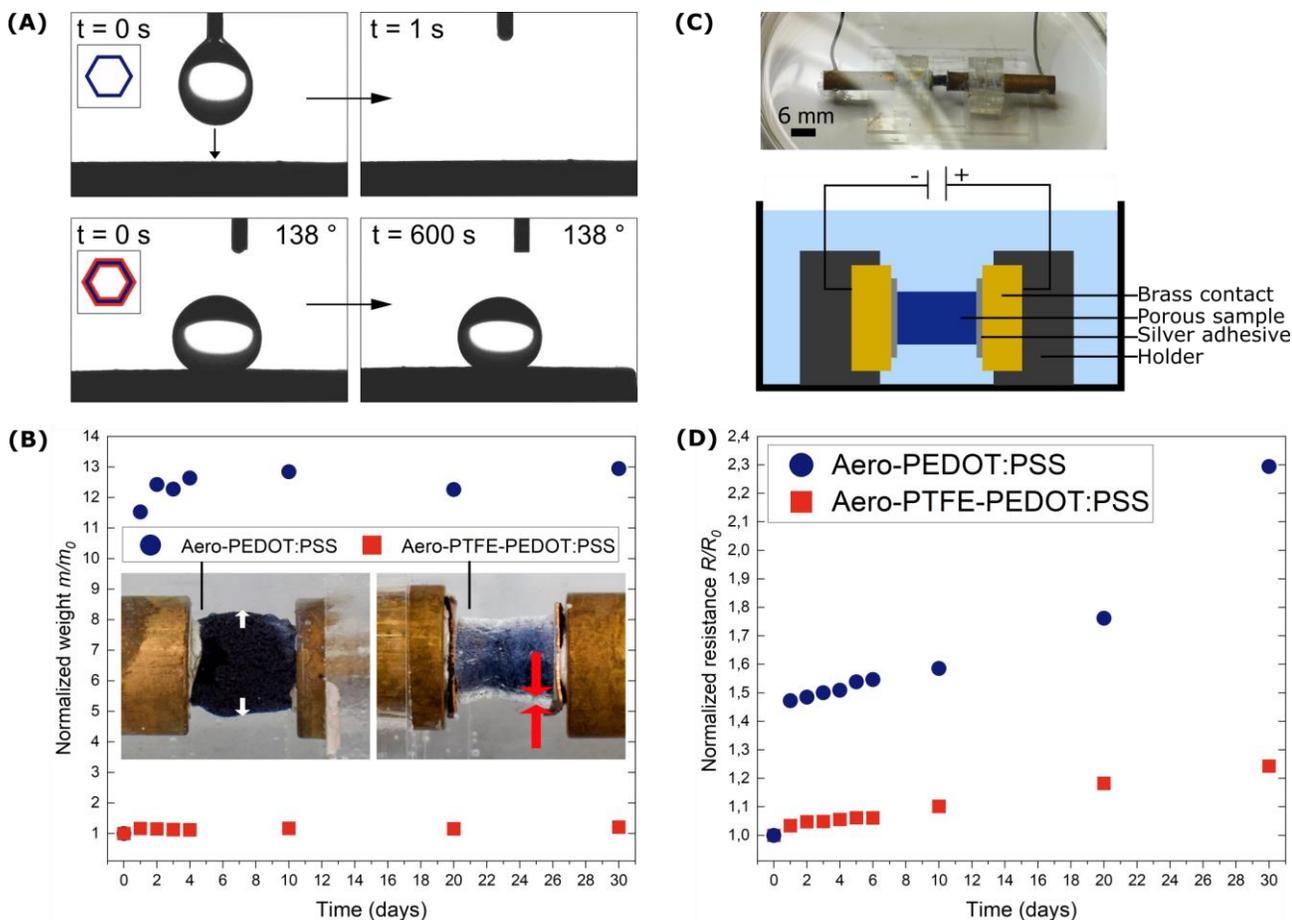

**Fig. 5. Water stability.** (**A**) Water contact angle measurement for Aero-PEDOT:PSS (top) and Aero-PTFE-PEDOT:PSS (bottom). Aero-PEDOT:PSS absorbs the droplet immediately, whereas the functionalized material shows a stable contact angle of about 138°. (**B**) Swelling ratios of Aero-PEDOT:PSS and Aero-PEDOT:PSS-PTFE after immersion in water for up to 30 days. The photographs indicate the swelling behavior. While bulging due to swelling is observed for Aero-PEDOT:PSS (white arrows), for Aero-PTFE-PEDOT:PSS a distinctive layer of air forms under water (red arrows), which highlights the strong hydrophobicity. (**C**) Experimental setup for IV-measurement immersed in water (bottom) and photograph of the setup (top). The porous samples are immersed in deionized water after being contacted with two metal electrodes. (**D**) Normalized resistance $R/R_0$ for Aero-PEDOT:PSS and Aero-PTFE-PEDOT:PSS after 30 days in water highlighting the greatly improved water stability of Aero-PEDOT:PSS.

## DISCUSSION

In summary, we demonstrate a new concept for the development of an ultralight, strain-invariant and highly water stable soft conductor. The concept is based on the coating of piezo-resistive microtubular framework structures by a nanoscopic insulating polymer layer, by which the macroscopic properties of the framework structure, such as mechanical, electrical, as well as wetting characteristics, can be tailored in a controlled manner. We specifically demonstrate, that ultra-light weight (< 15 mg/cm³), piezo-resistive PEDOT:PSS framework structures with a conductivity of up to 184 S/m, can be converted into strain-invariant soft conductors with constant conductivity between 80 % compressive and 25 % tensile strain by coating them with a nanoscopic insulating thin film (~ 23 nm) of Polytetrafluorethylene (PTFE). While the coating results only in a negligible weight increase, it enhances the mechanical flexibility and stability of the PEDOT:PSS framework structure, additionally providing highly hydrophobic properties. In combination, these properties overcome previous limitations of strain-invariant soft conductors with stable electrical properties, resulting in an attractive functional material system for application in the fields of soft robotics, wearable electronics, as well as electronic tissue and implants. While in this study, we focus on strain-invariant all organic soft conductors based on PEDOT:PSS and PTFE, this concept can be further transferred to tailor the properties of other ultra-lightweight framework materials and open porous aerogels, e.g. based on carbon or related nanomaterials, which have already demonstrated huge potential in a wide variety of applications, such as energy conversion, energy storage, thermal management, sensing, catalysis, water and air purification, as well as biomedicine. In summary, the here demonstrated concept represents thus a versatile approach to develop innovative functional soft materials and composites, by altering the macroscopic and microscopic properties of porous materials by a nanoscopic thin film coating process in 3D.

# MATERIALS AND METHODS

## Synthesis

*Preparation of ZnO networks:* ZnO tetrapods were prepared by flame transport synthesis, which has been described in great detail elsewhere (*43*, *44*). Sacrificial 3D networks of ZnO were prepared by mold forming ZnO tetrapods powder into the desired geometry. After subsequent sintering for 5 h at 1150°C in ambient conditions mechanically stable, interconnected ZnO networks were obtained.

*Preparation of PEDOT:PSS solutions*: Prior to usage PEDOT:PSS solution (Heraeus Clevios™ PH1000) was tip sonicated and filtered with a PES syringe filter (0.45 μm pore size). Aqueous solutions between $1 \cdot 10^{-6}$ to 0.6 wt% PEDOT:PSS were made by mixing a given volume of the pristine solution with deionized water and 5% v/v ethylene glycole, 3% v/v EtOH and 5% v/v divinyl sulfone (DVS) as crosslinking reagent. After heavy stirring for 3 minutes the solutions were immediately used. The solutions were kept at in a fridge in between infiltration cycles. For each new sample series fresh solution was prepared.

*Synthesis of free-standing 3D PEDOT:PSS networks:* The deposition of PEDOT:PSS onto the ZnO arms was carried out by dripping the freshly prepared PEDOT:PSS solution onto the template until the respective free volume was filled. After evaporation of the solvent and initiation of the crosslinking reaction on a hot-plate (4 h, 55 °C, air) the infiltration process was repeated multiple times. After completion of all infiltration and drying cycles, the networks were completely immersed in aqueous acid solution of 10M HCl for 24h. After repeated rinsing with deionized $H_2O$ and pure EtOH, the samples were dried by critical point drying (EMS 3000).

*iCVD coating of Aero-PEDOT:PSS:* The fabricated Aero-PEDOT:PSS samples were coated with a PTFE thin film via initiated chemical vapor deposition (iCVD). The monomer HFPO was combined with the initiator PFBSF for this purpose. Details on the iCVD reactor used can be found in the literature (*25*). The samples were coated in a continuous flow process. The substrate temperature of 20 °C was maintained by a thermostat (CC-K6, Huber) and a power of 72 Watt was applied to the NiCr filament (Goodfellow) array during the deposition by a power supply (Polaris 65-10, Knürr-Heinzinger). The process pressure of 50 Pa in the reactor was maintained by a butterfly valve (615, VAT), which receives feedback from a capacitive manometer (Baratron, MKS instruments) attached to the reactor. The vacuum was generated by a rotary vane pump (Duo 10, Pfeiffer Vacuum). The Monomer HFPO (0.5 sccm) was supplied to the reactor via a mass flow controller (MC, Alicat scientific) and the initiator PFBSF (0.1 sccm) was delivered to the reactor via needle valve (Varian).

## Characterization

*Conductivity measurements:* Cylindrical samples (height 6mm, diameter 6mm) were attached to a custom made electrode holder (see S1) via silver conductive paste. Afterwards linear voltage sweeps using a Keithley 2401 in 4 wire sense mode were performed. The conductivity for each sample was calculated from averaging the slope from 10 sweeps and 3 different samples.

*Electromechanical characterization:* Cylindrical samples (6mm height, 6mm diameter) were contacted to a custom made electromechanical characterization setup (Märzhäuser Wetzlar HS 6.3 micromanipulator, burster 9235/36 force sensor) via silver conductive paste (see Fig. S9). All measurements were performed at a voltage of 0.5 V.

*Fourier transform infrared spectroscopy (FTIR)*: A solution with 0.25 wt. % PEDOT:PSS was used to deposit thin films via spin-coating (Laurell WS-650-23B) on 20x20 mm Silicon substrates. Prior to deposition the Si substrates were immersed in Piranha solution to increase hydrophilicity. After spin-coating at 400 rpm for one minute, the samples were dried at 65 °C on a hot plate for 6 hours. The resulting film thickness of ~400 nm was measured by profilometry (Bruker DektakXT). PTFE functionalization was performed via iCVD by depositing for 30 minutes at 72 W with flows for PFBSF of 0.1 sccm and HFPO of 0.5 sccm. FT-IR spectra of functionalized and non-functionalized samples were obtained in a range from 4000 to 500 cm$^{-1}$ in air (Bruker INVENIO). Background correction and peak analysis was performed using Origin 2022 software.

*X-ray photoelectron spectroscopy (XPS):* For XPS analysis (Omicron Nanotechnology GmbH, Al Kα X-ray anode, 240 W power) aero-samples with a geometry of 10x10x2 mm samples were prepared. Prior to analysis a thin strip of 100 nm gold was sputter-deposited on the edge of the samples for binding energy reference. For binding energy calibration the Au $4f_{7/2}$ peak (84.00 eV) was used as reference value. Peak analysis was performed using the CasaXPS software.

*Electrical characterization underwater:* Cylindrical samples (6mm height, 6mm diameter) were contacted to a custom made electromechanical characterization setup (Fig. 5C). At first the UI-curves were measured using a constant voltage of 0.5 V for 1000 minutes. Afterwards linear voltage sweeps from -0.5 to +0.5 V were performed for each separate day from which the resulting resistance has been calculated.

*Contact angle measurement:* For contact angle measurements samples with a geometry of 10x10x2 mm were fabricated. Static contact angle measurements (OCA 30, Dataphysics) were performed by depositing 10 µl droplets of deionized water onto the surface.

*Swelling behavior*: For the estimation of water uptake cylindrical samples (6 mm height, 6 mm diameter) were immersed in deionized water. Aero-PTFE-PEDOT:PSS were held under water by a sieve structure. After the respective amount of days, samples were removed from the water, put on tissue to remove excess water for a fixed amount of time (1 minutes) and their weight was measured using a microscale (Sartorius Micro MC1). Swelling ratios were calculated by comparison to the initial weight in dry state.

*Microstructure*: An SEM Ultra 55 (Zeiss) has been used to perform scanning electron microscopy at a scanning distance of 3 mm and acceleration voltage of 3 kV. Samples were glued to custom made holders prior to measurements.

*Transmission electron microscopy (TEM):* TEM analysis has been performed with a FEI Tecnai F30 STwin operated at 300 kV. Samples were prepared for TEM analysis by gentle grinding a small portion of a template submerged in n-Butanol in a mortar. The resulting suspension was drop coated onto lacey carbon coated copper TEM grids. After drying in air single microrods are readily available for analysis. By simultaneously utilizing material of all areas from a cross sectional template a random selection of microrods is achieved. Film thicknesses have been estimated by taking 4-7 measurement points at 12 seperate locations for each sample.

**Supplementary Materials**
Figs. S1 to S9
Tables S1 to S4
Movies S1 to S6

**Acknowledgments:**

**Funding:**
German Research Foundation grant GRK 2154 (IB, FS, FR, RA)
German Research Foundation grant SCHU 3506/4-1 (FS, RA)
German Research Foundation grant SFB 1461 – Project-ID 434434223 (RA)
German Research Foundation grant SFB1261 Project A2 (SS, TS, FF)
German Research Foundation grant KI 1263/17-1 (NK, LK)

**Author contributions:**
Examples:
Conceptualization: FS, SS
Methodology: IB, FS, SS, FR
Investigation: IB, NK
Visualization: IB, FS, SS, NK
Funding acquisition: LK, FF, RA, FS
Project administration: LK, FF, RA, FS
Supervision: FS, SS
Writing – original draft: IB, NK, FS, SS
Writing – review & editing: TS, LK, FF, RA, FS, SS, FR

**Competing interests:** Authors declare that they have no competing interests.

**Data and materials availability:** All data are available in the main text or the supplementary materials

# Supplementary Materials for

## Strain-invariant, highly water stable all-organic soft conductors based on ultralight multi-layered frameworks structures

**Authors:**

I. Barg,[1] N. Kohlmann,[2] F. Rasch,[3] T. Strunskus,[1,4] R. Adelung,[3,4] L. Kienle,[2,4] F. Faupel[1,4], S. Schröder, [1,4]*†
F. Schütt[3,4]* †

**Affiliations:**

[1]Chair for Multicomponent Materials, Institute for Materials Science, Kiel University; 24143 Kiel, Germany
[2]Chair for Synthesis and Real Structure, Institute for Materials Science; 24143 Kiel University, Kiel, Germany
[3]Chair for Functional Nanomaterials, Institute for Materials Science, Kiel University; 24143 Kiel, Germany
[4]Kiel Nano, Surface and Interface Science KiNSIS, Kiel University; 24118 Kiel, Germany

*Corresponding authors. Email:
Dr.-Ing. Fabian Schütt (fas@tf.uni-kiel.de)
Dr.-Ing. Stefan Schröder (ssch@tf.uni-kiel.de)

†Shared senior authorship

**This PDF file includes:**

Figs. S1 to S9
Tables S1 to S4
Movies S1 to S5

1. <u>Comparison tables for strain-invariance and fatigue</u>

**Table S1.**

Comparison of key values regarding mechanical stability during long-term cycling of porous and Aeromaterials.

| Ref. | Material | Energy loss coefficient (Start→end) | Cycles | Strain / % | Residual strain / % | Residual normalized max. stress $\sigma/\sigma_0$ |
|---|---|---|---|---|---|---|
| This work | Aero-PTFE-PEDOT:PSS | 0.28→0.27 | 2000 | 50 | 7 | 0.98 |
| (13) | PEDOT:PSS-BIBSAT foam | / | 1000 | 40 | / | 0.71 |
| (19) | Polyimide Aerogel | 0.48→0.28 | 10000 | 50 | 5.80 | 0.82 |
| (21) | Boron nitride aerogel | 0.62→0.49 | 2000 | 50 | 12 | 0.85 |
| (45) | Mxene/Polyimide Aerogel | 0.4→0.32 | 1000 | 50 | 7 | 0.88 |
| (46) | Carbonaceous nanofibrous aerogels | 0.48→0.33 | 1000 | 50 | 4.3 | 0.75 |
| (47) | Aramid nanofibre/polyimide aerogel | 0.64→0.43 | 1000 | 60 | 12 | 0.85 |
| (48) | C-G monolith | 0.29→0.19 | 250000 | 50 | 2 | 0.86 |
| (49) | Graphene aerogel | 0.75→0.62 | 1000 | 50 | 10 | 0.88 |
| (50) | Graphene cellular network | 0.85→0.55 | 10 | 50 | 5 | 0.68 |
| (51) | Ni microlattice | 0.8→0.42 | 10 | 50 | 3 | 0.7 |
| (52) | Graphene-CNT aerogels | 0.8→0.43 | 1000 | 50 | 10 | 0.88 |
| (53) | PP aerogel / PAAm hydrogel hybrid | / | 5*10^5 | 40 | 5.2 | 0.91 |
| (54) | Silicia nanofibrous aerogel | 0.77→0.43 | 1000 | 60 | 6 | 0.7 |
| (55) | Graphene aerogel | 0.7→0.5 | 10^5 | 70 | / | 0.7 |
| (56) | $C_2T_x$ Mxene/Acidified CNT Aerogel | / | 100 | 50 | 7 | 0.79 |
| (57) | PI-Nanofibre/Mxene aerogel | 0.3→0.2 | 1000 | 50 | / | 0.97 |
| (58) | Carbon aerogel | / | 500 | 50 | 0.04 | 0.81 |
| (59) | Carbon aerogel | / | 100 | 50 | 5 | 0.88 |
| (60) | PI / CNT aerogel | 0.45->0.2 | 1000 | 30 | 7 | 0.99 |

**Table S2.**

Overview on key values regarding the electrical stability during single and long-term cycling of soft conductors.

| Ref. | Material | Single load performance $R/R_0$ at max strain* | Strain / % | Long-term cyclic performance $\Delta R/R_0$** | Strain / % | Cycles |
|---|---|---|---|---|---|---|
| This work | Aero-PTFE-PEDOT:PSS | 0.95 | 80 | +0.0 | 50 | 2000 |
| (13) | PEDOT:PSS-BIBSAT aerofoam | 1.1/0.7 | +60,-60 | +0.0 | -40,+20 | 1000 |
|  |  |  |  | +0.45 | -60,+40 | 1000 |
| (42) | PEDOT:PSS/Polyacrylamide nanoweb | / | / | +0.6 | 50 | 2000 |
|  |  |  |  | +3 | 50 | 10000 |
| (53) | PP aerogel / PAAm hydrogel hybrid | 1.5 | 5 | / | / | / |
| (58) | Carbon Aerogel | 0.15 | 50 | +0.1 | 50 | 10 |
| (59) | Carbon aerogel | 0.1 | 60 | / | / | / |
| (60) | PI / CNT aerogel | 0.1 | 80 | +0.09 | 30 | 1000 |
| (61) | PEDOT:PSS + D-Sorbitol thin film | 1.3 | 60 | +0.9 | 60 | 10 |
| (62) | PEDOT:PSS + D-Sorbitol + Waterborne Polyurethane | 1.1 | 30 | +0.09 | 30 | 400 |
| (63) | PEDOT:PSS + ionic liquids thin film | 0.36 | 50 | +1.1 | 50 | 1000 |
| (64) | Au/PDMS thin film | 1.5 | 50 | +0.5 | 50 | 500 |
| (65) | PEDOT:PSS + PAAMPSA thin film | 1.25 | 60 | / | / | / |
| (66) | Ag NP + elastomeric fibres | 1.59 | 80 | / | / | / |
| (67) | MWCNT/RGO @ Polyurethane aerosponges | 1.59 | 60 | / | / | / |
| (68) | Au Ag Cu @ PDMS Aeronetwork | / | / | +3.0 | / | 2000 |
| (69) | Fluorinated Graphene Aerogel | 0.09 | 40 | / | / | / |
| (70) | PPy on porous PDMS | 0.8 | 50 | / | / | / |
| (71) | Pt NP on Porous PDMS foam | 20 | 50 | / | / | / |
| (72) | Vertically aligned CNTS in PU/PDMS | 1.53 | 50 | / | / | / |
| (73) | PU + CuAg + PDMS sponge | / | / | +1.2 | 40 | 1000 |
| (74) | Joint-welded CNT foam | 0.92 | 80 | / | / | / |
| (75) | PEDOT:PSS-PAAm organogel | / | / | +0.56 | 50 | 1000 |
| (76) | MWCNT @ PU | / | / | -0.7 | 180 | 5 |

\* For a single load $'R/R_0$ at max. strain' denotes the resistance at the respective strain normalized with respect to the resistance at zero strain.

\*\* For long-term cyclic performance '$\Delta R/R_0$' denotes the change in the resistance at zero strain after the last cycle normalized with respect to the resistance at zero strain at the first cycle

**Table S3.**

Overview on key values regarding the electrical stability during single and long-term cycling of soft conductors.

| Ref. | Material | Water contact angle / ° | Long term measurement | |
|---|---|---|---|---|
| | | | ΔR/R0 | Method |
| This work | Aero-PTFE-PEDOT:PSS | 138 -> 138 (10 minutes) | 1.24 | 30 days immersion in water |
| (13) | PEDOT:PSS-BIBSAT aerofoam | / | 4.0 | 20 minutes immersion in DI water |
| (41) | PEDOT:PSS in PDMS matrix | 121 -> 109 (8 seconds) | 'stable' | 20 minutes immersion in DI water |
| (42) | PEDOT:PSS/Polyacrylamide nanoweb | / | 1.27 | 2 h immersion in DI water |
| (77) | PEDOT:PSS-PBA/silica fluorinated thin film | / | 1.43 | 90 % RH at 30 °C for 30 days |
| (78) | PEDOT in PU matrix | / | 1.4 | 50 % RH at 60 °C for 17 hours |

2. Conductivity measurements and conductivity enhancement

To measure the specific conductivity samples (6mm height, 6mm diameter) were contacted on custom made holders by silver conductive paste (Fig. S1). The IV-characteristics were measured with 10 linear sweeps from -0.5 V to + 0.5 V in 4 wire sense mode using a Keithley 2401. After estimating the average resistance $R$ the respective conductivity $\sigma$ was calculated using:

$$\sigma = \frac{1 \cdot h}{R \cdot \pi \cdot r^2}$$

where $\sigma$ is the conductivity, $h$ is the height of the template, and $r$ is the radius of the sample.

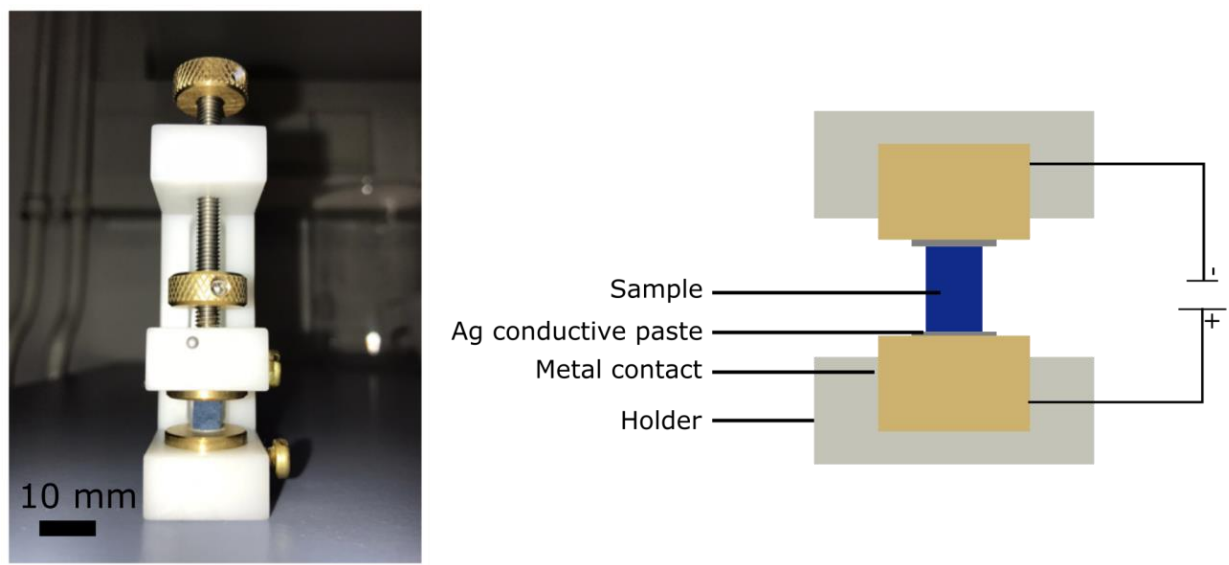

**Fig. S1.**

Custom made holders used for conductivity measurements with PEDOT:PSS-ZnO sample loaded and corresponding schematic drawing.

The mechanism of conductivity increase has been investigated by XPS measurements of PEDOT:PSS-ZnO before etching and Aero-PEDOT:PSS after etching. Both the conductive PEDOT and the non-conductive PSS have a sulfur compound with thiophene for PEDOT and sulfonate for PSS. Hence the analysis is based on the calculation of the area ratio between the respective S 2p peaks. S 2p core-levels consist of spin-spin doubles (S $2p_{1/2}$ and S $2p_{3/2}$) with an energy splitting of 1.18 eV, same FHWM and relative intensity ratio (1:2). For the sulfonate group of PSS two variations can be present: neutral PSSH and at a slightly lower binding energy negatively charged $PSS^-$. The binding energy difference is $\Delta E = 0.4$ eV. After a Tougaard type background correction, a mixture of Gaussian and Lorentzian peak shapes was used for peak fitting. For PEDOT an asymmetric peak shape with a high binding energy tail was used. The asymmetry corresponds to partially p-doped PEDOT regimes with delocalized positive charges over multiple chains, for which the binding energy depends on the local distance to negatively charged PSS-counterions.

The S 2p peak analysis and deconvolution is given for PEDOT:PSS-ZnO in Figure S5a and Aero-PEDOT:PSS in Figure S5b. The PEDOT S 2p peak can be found at S $2p_{1/2}$ = 165.0 eV and S $2p_{3/2}$ = 163.8 eV. The positive PSSH (S $2p_{1/2}$ = 168.5 eV and S $2p_{3/2}$ = 169.7 eV) is present at a slightly higher binding energy than the negatively charged PSS- (S $2p_{1/2}$ = 168.0 eV and S $2p_{3/2}$ = 169.2 eV). The ratio between neutral PSSH and PSS- changes after etching, with a higher amount of PSS- present. The suggested mechanism is the fragmentation and subsequent removal of neutral PSSH during the etching process (*79*)(*80*). Since both PSSH and PSS- area non-conductive, they were treated interchangeably in the calculation between conductive PEDOT and non-conductive PSS. The calculated area ratio between PSS to PEDOT is 3.51 for PEDOT:PSS-ZnO and 1.48 for Aero-PEDOT:PSS. Thus the PSS content has been reduced by 56.6 %.

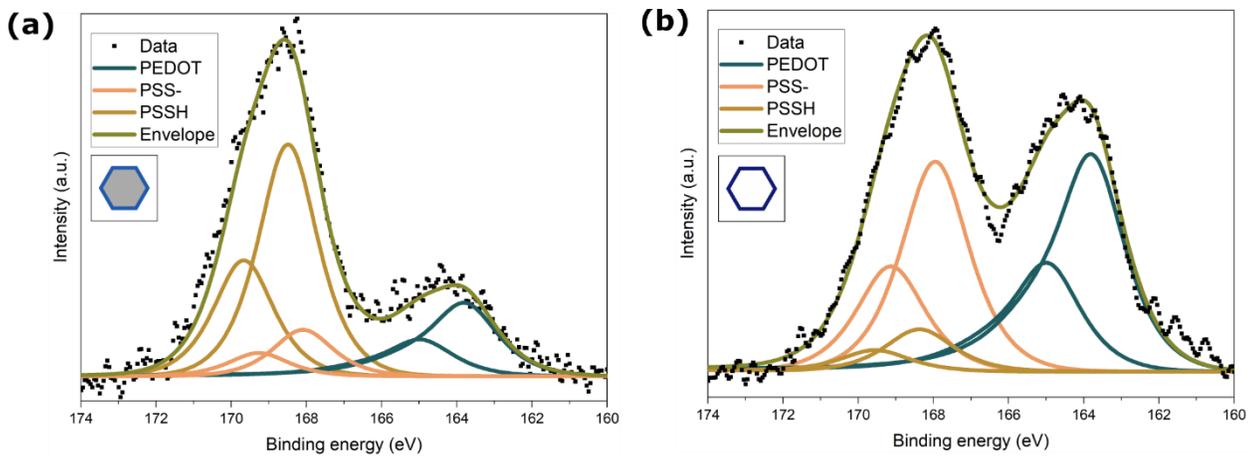

**Fig. S2.**
Deconvolution of Sulfur 2p peaks for **a)** PEDOT:PSS-ZnO composite before etching and **b)** Aero-PEDOT:PSS after HCl etch removal of ZnO.

3. <u>Morphology investigation by scanning electron microscopy</u>

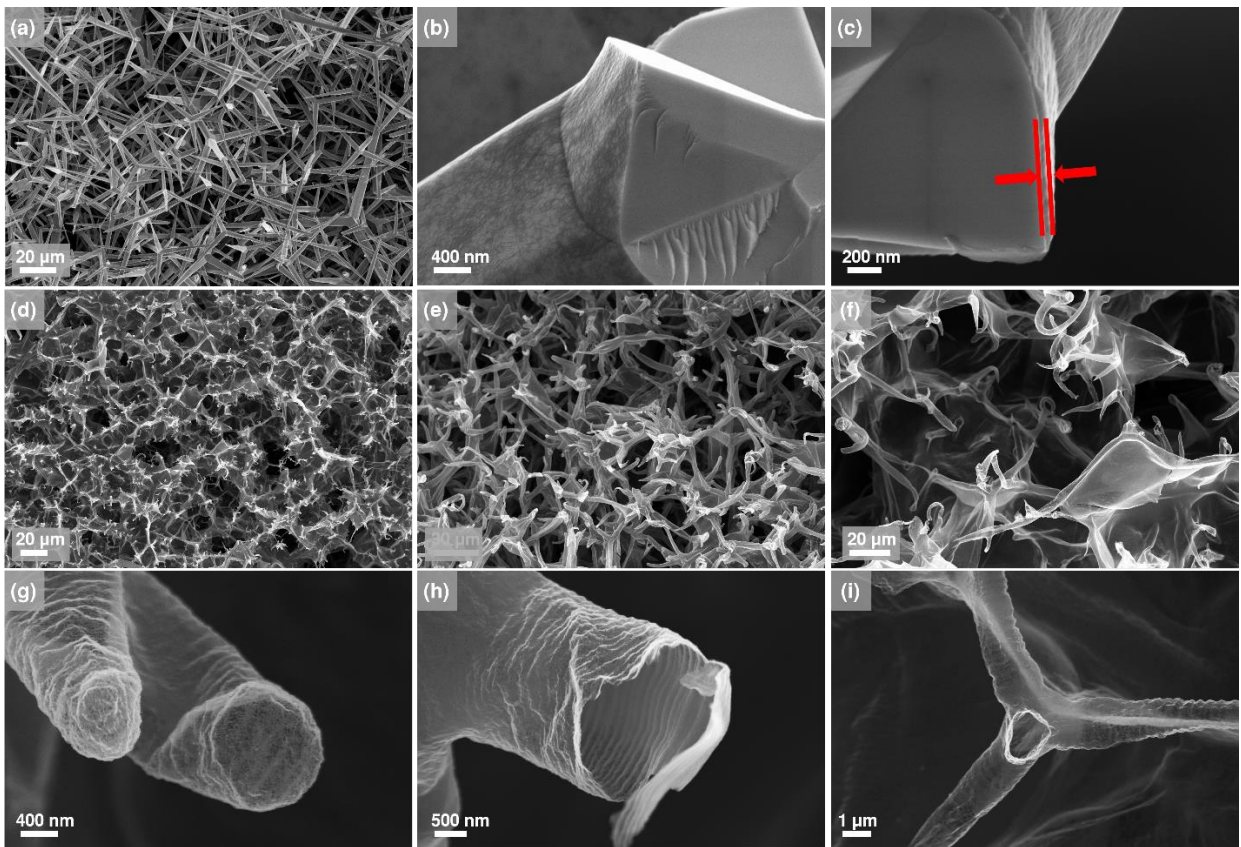

**Fig. S3.**
SEM micrographs highlighting the morphology of t-ZnO, PEDOT:PSS on t-ZnO and freestanding PEDOT:PSS. **a)** Pristine t-ZnO network after sintering with a density of 0.3 g/cm³. **b)** Partially broken ZnO arm after infiltration with PEDOT:PSS layer on the surface. **c)** Side-view of partially broken ZnO arm with view on a PEDOT:PSS layer indicated by the red markers. **d)** Image of PEDOT:PSS sample surface after etch removal of ZnO. **e)** Overview image of PEDOT:PSS network after etch removal highlighting that the tetrapod shape is kept intact. **f)** PEDOT:PSS exhibits sail structures in between the arms in some areas of the sample. **g)** Free-standing arm with PEDOT:PSS shell indicates the high porosity and hollow nature of the material. The porosity of the PEDOT:PSS arms can be attributed to removal of PSS by etching. **h)** Inside view into a hollow arm of PEDOT:PSS which resembles the structure of the previously present ZnO arm. **i)** Top view on hollow PEDOT:PSS tetrapod arm with one broken of arm.

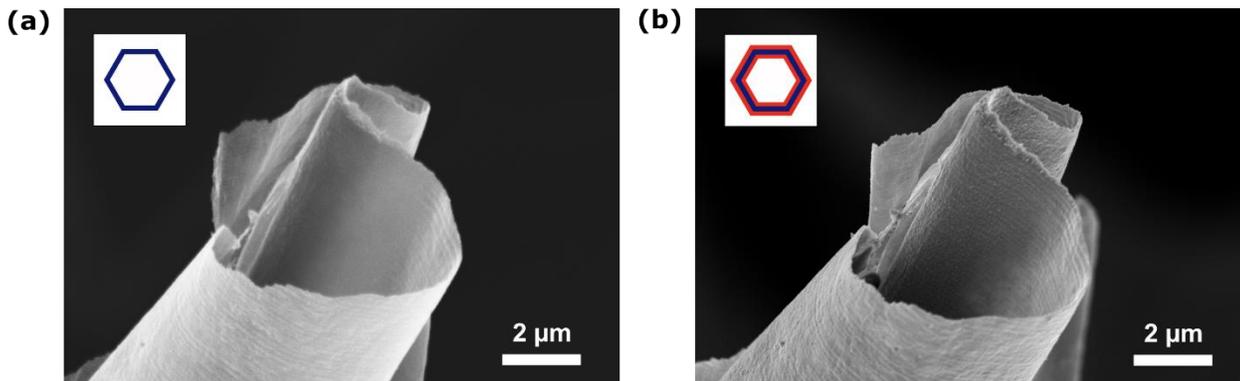

**Fig. S4.**
SEM micrographs of the same Aero-PEDOT:PSS structure highlighting the morphology of **a)** an Aero-PEDOT:PSS arm before and **b)** an Aero-PTFE-PEDOT:PSS arm after functionalization via iCVD. In SEM no immediate changes in morphology stand out.

4. <u>Fourier-transform infrared spectroscopy</u>

**Table S4.**

Peak positions and their corresponding bonding of PEDOT and PSS for PEDOT:PSS and PTFE-PEDOT:PSS obtained through FTIR.

| Position / cm$^{-1}$ | Bond | Component |
|---|---|---|
| 1583 | C=C | PSS |
| 1537 | C=C | PEDOT |
| 1308 | C-C | PEDOT |
| 1220 | C-O-C | PEDOT |
| 1135 | C-O-C | PSS |
| 1083 | C-O-C | PEDOT |
| 1030 | S-O | PSS |
| 972 | C-S | PEDOT |
| 920 | C-S | PEDOT |
| 831 | C-S | PEDOT |
| 685 | C-S | PSS |

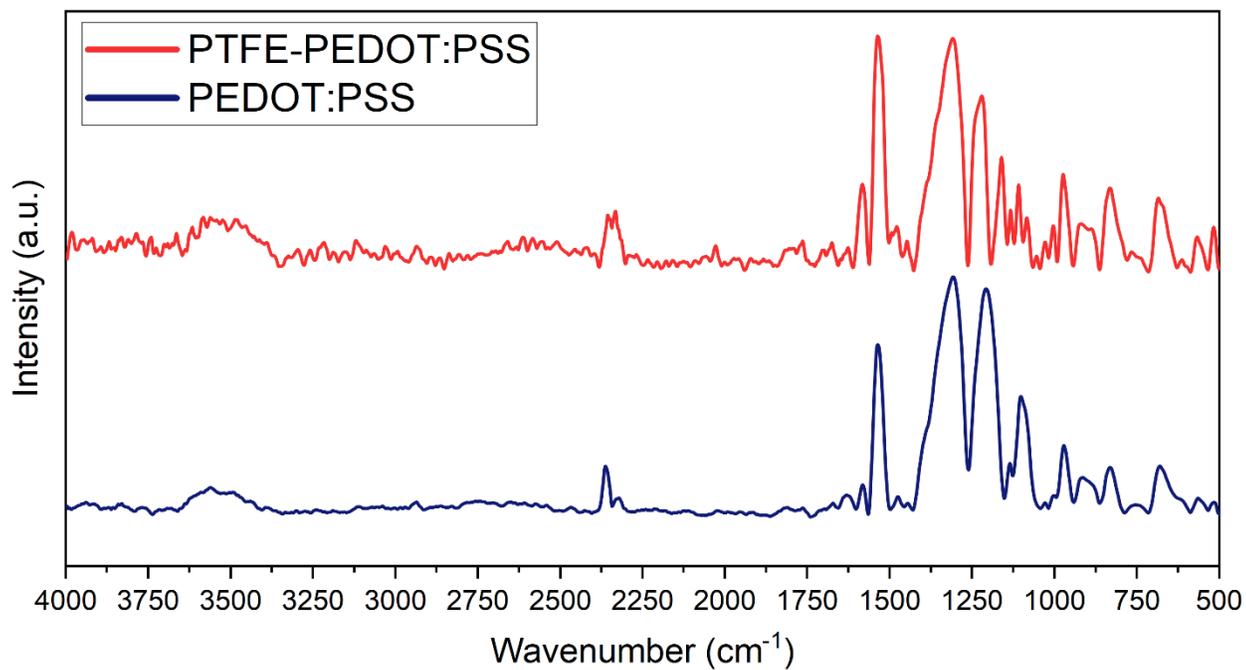

**Fig. S5.**
FTIR spectra between 4000 and 500 cm$^{-1}$ for PTFE-PEDOT:PSS and PEDOT:PSS. The peaks at around 2350 cm$^{-1}$ are related to $CO_2$. The broad peak region around 3500 cm$^{-1}$ is associated with O-H stretching vibrations.

5. X-ray photoelectron spectroscopy

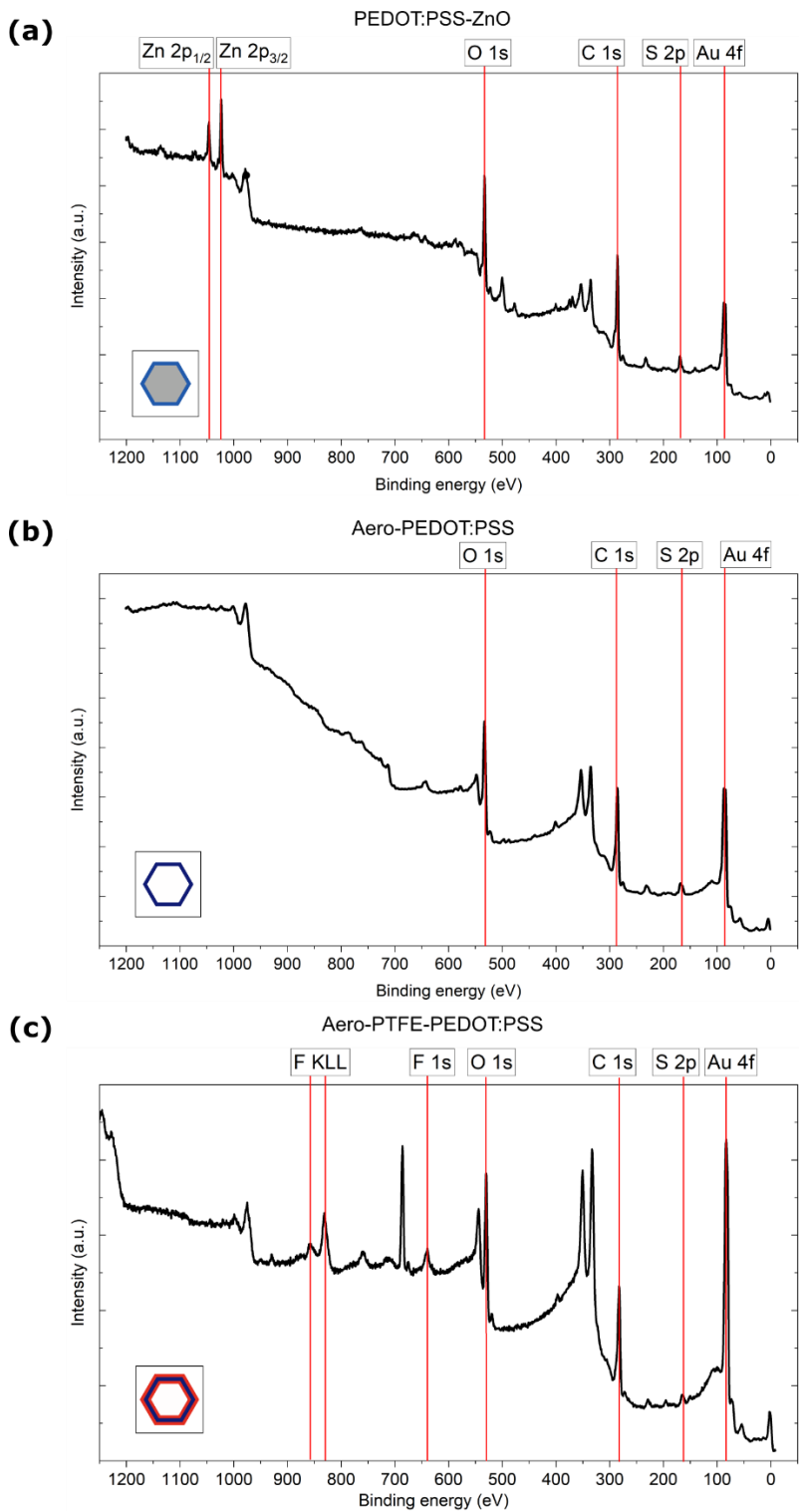

**Fig. S6.**
Wide XPS-Spectra for different samples. **a)** PEDOT:PSS-ZnO before etch removal ZnO with corresponding Zn 2p peaks. **b)** Aero-PEDOT:PSS after etch-removal by immersion in HCl for 24h and critical point drying. **c)** Aero-PTFE-PEDOT:PSS with characteristic F 1s peak at 688.9 eV and F KLL peaks at 833.5 eV and 859.5 eV.

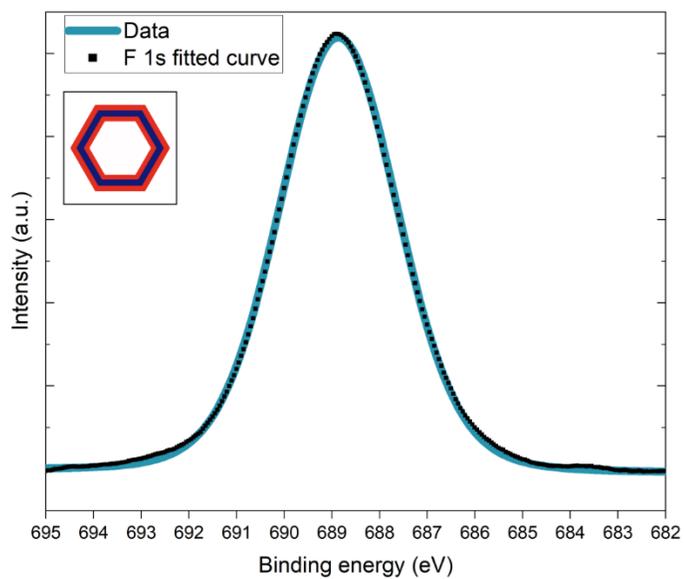

**Fig. S7.**
High resolution scan and deconvolution Fluorine 1s peak for Aero-PTFE-PEDOT:PSS. The appearance of the peak at 688.9 eV indicates the succesful deposition of PTFE and is well in agreement with reported values (*81*).

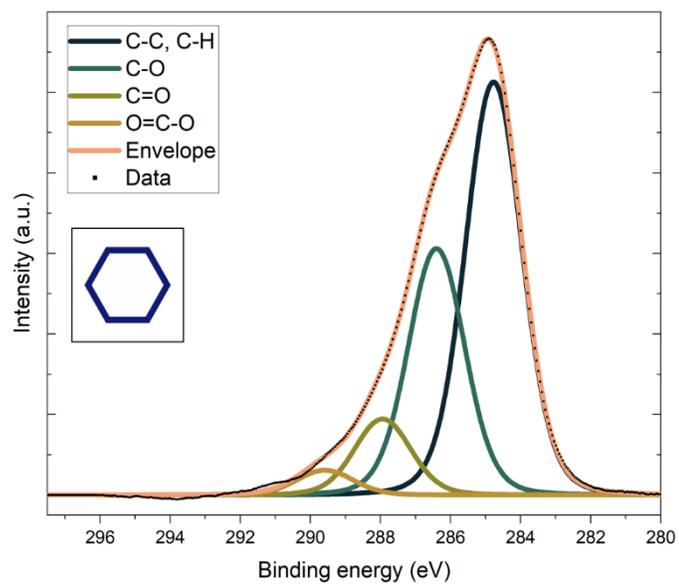

**Fig. S8.**

High resolution scan and deconvolution of C 1s peak for PEDOT:PSS-ZnO before etching and subsequent functionalization

6. <u>Mechanical and electrical characterization setup</u>

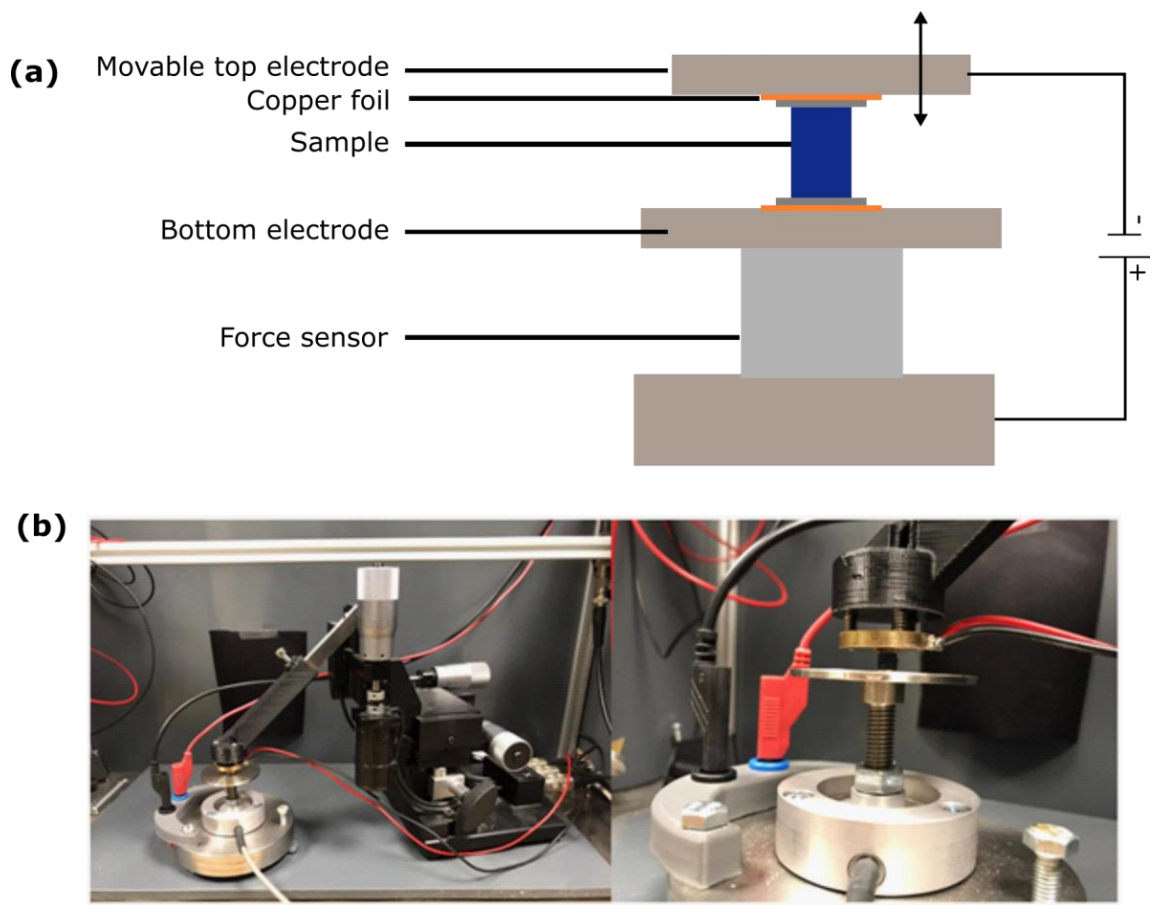

**Fig. S9.**
**a)** Schematic overview of electromechanical characterization setup. The samples are attached to the metal electrodes via silver conductive paste. A micromanipulator moves the top electrode while the resulting force is picked up by a screw-force sensor. **b)** Photograph showing the measurement setup.

7. <u>Calculation of elastic energy</u>

The mechanical energy $U$ per half-cycle was calculated via the area under the loading/unloading curve:
$$U = \int_{\varepsilon}^{\varepsilon_{max}} F(\varepsilon)\, d\varepsilon$$
with $\varepsilon$ being the strain at the intersection with the x-axis, $\varepsilon_{max}$ the strain at maximum elongation and $F$ the force at the respective position.

The energy loss coefficient is then calculated as normalized ratio difference between the loading curve $U_{load}$ and unloading curve $U_{unload}$:
$$Q = \frac{U_{load} - U_{unload}}{U_{load}}$$

**Movie S1.**

Wet-chemical infiltration of ZnO template with a PEDOT:PSS solution.

**Movie S2.**

Compression up to 80 % for Aero-PTFE-PEDOT:PSS (left) and Aero-PEDOT:PSS (right).

**Movie S3.**

Wetting behavior during contact angle measurements for Aero-PEDOT:PSS highlighting the immediate water absorption.

**Movie S4.**

Wetting behavior of Aero-PTFE-PEDOT:PSS highlighting the high water repellency of the material.

**Movie S5.**

The outstanding hydrophobicity combined with the low density is highlighted as Aero-PTFE-PEDOT:PSS can't be submerged in water.